\documentclass{article}
\usepackage{young3}
\usepackage{graphicx}

\begin{document}
\atitle {A distribution of the temperature in a ring of the
incompressible, viscous liquid with two free boundaries. \\Exact
solutions.}{Belmetsev~N.F., Bytev~V.O.} [Belmetsev~N.F.,
Bytev~V.O. A distribution of the temperature in a ring of the
incompressible, viscous liquid with two free boundaries. \\Exact
solutions.]
\vspace{-1.6\baselineskip} \centerline{Tyumen State
University, Russia} \vspace{\baselineskip}

\vspace{\baselineskip}

\vspace{-1.75\baselineskip} \centerline{\it e-mail:\
weqsmachine@gmail.com, vbytev@utmn.ru} \vspace{\baselineskip}
\centerline{October, 2009} \vspace{\baselineskip}

\textbf{Abstract.} It has been found the exact solutions for
nonstationary distribution of the temperature in the liquid ring
with two viscosities ($\mu$ and $\mu_0$) and two free boundaries
of the ring. \vspace{\baselineskip}

We consider the problem about rotational symmetric movement of
viscous and nonviscous liquid in a ring for nonclassical model of
hydrodynamics (see \cite{ANDREEV1B, BYTEV_B}). A similar problem
for classical models Navier-Stokes and Stokes was considered in
papers {\cite{OVSYANNIKOV1B, BYTEV1B, PUHNACHEV1B, LAVRENTEVA1B,
LAVRENTEVA2B}}. Previous papers contain the proves for theorems of
global resolvability and uniqueness, and contain the asymptotes
for solutions behavior. In the moment $t = 0$, we see that the
liquid fills a ring $R_{20} < r < R_{10}$ and has distribution of
velocities: $$u_r  = \varphi _0 \,r^{ - 1},\qquad u_\theta =
\upsilon _0 (r).$$ Here, $R_{10}$, $R_{20}  > 0$, and $\varphi _0$
are constants, $\upsilon _0 (r)$ is an arbitrary smooth function.
Since $u_\theta$ is not depend from $\theta$, we see that the
specified association $u_r$ from $r$ is unique.

Let $R_{10}$, $R_{20}$ are free boundaries. In this case the
liquid ring begins movement by inertia. Since we have rotational
symmetry of an initial condition, we may assume that this one is
similarly in all time agreement. Using this statement, we get that
free boundaries are remained circles during all time of movement:
$r = R_1 (t)$, $r = R_2(t)$.

Let us note the equations of movement and thermal conduction for
nonclassical liquid \cite{ANDREEV1B, BYTEV_B} with the supposition
of its incompressibility and homogeneity:

\begin{equation*}
 \frac{{\partial \vec u}}{{\partial t}} + (\vec u \cdot \nabla
 )\vec u - \frac{1}{\rho }M\Delta \vec u + \frac{1}{\rho }\nabla p = 0
\end{equation*}
\begin{equation*}
 div\,\vec u = 0
\end{equation*}
\begin{equation*}
 \rho \,C_p \frac{{dT}}{{dt}} = k\,\Delta T + \rho \,\nu \,H
\end{equation*}
Here, $\vec u$ is a vector of velocity with components $u_i$; $p$
is a hydrostatic (equilibrium) pressure; $\rho$ is a density; $H$
is a dissipative function; $C_p$ is a thermal capacity; $k$ is a
factor of thermal conduction; $\mu = \rho \,\nu$ is a viscosity;
$M$ is a tensor of viscosity:
\begin{equation*}
 M = \left( {\begin{array}{*{20}c}
   \mu  & {\mu _0 }  \\
   { - \mu _0 } & \mu   \\
\end{array}} \right)
\end{equation*}
$\mu _0$ is nondissipative viscosity which can accept values with
any signs. The stress tensor:
\begin{equation*}
    T =  - pI + 2MD
\end{equation*}
$D$ is the tensor with components:
\begin{equation*}
    D_{ij}  = \frac{1}{2}\left( {\frac{{\partial u_i }}{{\partial x_j
}} + \frac{{\partial u_j }}{{\partial x_i }}} \right),\qquad(i,j =
1,2).
\end{equation*}

Let $r = R_1 (t)$ be an exterior boundary of the ring and $r = R_2
(t)$ be an interior boundary of the ring. Since,
\begin{equation*}
    u_r  = u(r,t);\qquad u_\theta   = \upsilon(r,t);\qquad p =
    p(r,t),
\end{equation*}
then a components of the stress tensor in polar coordinates are:
\begin{eqnarray*}
  T_{rr}&=&- p + 2\mu \frac{{\partial u}}{{\partial r}} + \mu _0
\left( {\frac{{\partial \upsilon }}{{\partial r}} - \frac{\upsilon
}{r}} \right); \\
  T_{r\theta}&=&\mu \left( {\frac{{\partial
\upsilon }}{{\partial r}} - \frac{\upsilon }{r}} \right) - 2\mu _0
\frac{{\partial u}}{{\partial r}}; \\
  T_{\theta \theta}&=&- p +
2\mu \frac{u}{r} - \mu _0 \left( {\frac{{\partial \upsilon
}}{{\partial r}} - \frac{\upsilon }{r}} \right).
\end{eqnarray*}

Using this representation, we obtain:
$$
u = \frac{{\Phi (t)}}{r};$$
$$
\frac{1}{r}\frac{{d\,\Phi }}{{dt}} - \frac{{\Phi ^2 }}{{r^3 }} -
\frac{{\upsilon ^2 }}{r} =  - \frac{\partial }{{\partial r}}\left(
{p - \nu _0 \frac{1}{r}\frac{{\partial (r\upsilon )}}{{\partial
r}}} \right);$$
$$
 \frac{{\partial \upsilon }}{{\partial t}} +
\frac{\Phi }{r}\frac{{\partial \upsilon }}{{\partial r}} +
\frac{{\Phi \upsilon }}{{r^2 }} = \nu \left( {\frac{{\partial ^2
\upsilon }}{{\partial r^2 }} + \frac{1}{r}\frac{{\partial \upsilon
}}{{\partial r}} - \frac{\upsilon }{{r^2 }}} \right);$$
$$
 \rho \,C_p \left( {\frac{{\partial T}}{{\partial t}} +
u\frac{{\partial T}}{{\partial r}} + \frac{\upsilon
}{r}\frac{{\partial T}}{{\partial \theta }}} \right) = k\left[
{\frac{1}{r}\frac{\partial }{{\partial r}}\left( {r\frac{{\partial
T}}{{\partial r}}} \right) + \frac{1}{{r^2 }}\frac{{\partial ^2
T}}{{\partial \theta ^2 }}} \right] +{}$$
$$
{}+ \mu\left\{ {2\left( {\frac{{\partial u}}{{\partial r}}}
\right)^2  + 2\left( {\frac{u}{r}} \right)^2  + \left[
{r\frac{\partial }{{\partial r}}\left( {\frac{\upsilon }{r}}
\right)} \right]^2 } \right\}.$$ Here $u$ is the radial component
of velocity; $\upsilon$ is the azimuthal component of velocity.
Since the stress vector is equal to zero on free boundaries, we
get boundary conditions:
\[
T_{rr}  =  - \left( {p + \frac{{2\,\nu \,\Phi (t)}}{{r^2 }}}
\right) + \nu _0 \left( {\frac{{\partial \upsilon }}{{\partial r}}
- \frac{\upsilon }{r}} \right) = 0, \,\, \text{if} \,\, r =
R_{1,2}(t);
\]
\[
T_{r\theta }  = \nu \,r\frac{\partial }{{\partial r}}\left(
{\frac{\upsilon }{r}} \right) - 2\nu _0 \frac{{\Phi (t)}}{{r^2 }}
= 0, \,\, \text{if} \,\, r = R_{1,2}(t).
\]
Also, we have dynamic condition on free boundaries:
\[
\frac{{dR_i (t)}}{{dt}} = \frac{{\Phi (t)}}{{R_i (t)}},\,\,\,(i =
1,2)
\]

Supposing  $ \mu _0^2  + \mu ^2  \ne 0 $, we have:
\[
\frac{{\partial \upsilon }}{{\partial t}} + \frac{\Phi
}{r}\frac{{\partial \upsilon }}{{\partial r}} + \frac{{\Phi
\upsilon }}{{r^2 }} = \nu \frac{\partial }{{\partial r}}\left[
{\frac{1}{r}\frac{{\partial (r\upsilon )}}{{\partial r}}} \right];
\]
\[
\nu \left( {\frac{{\partial \upsilon }}{{\partial r}} -
\frac{\upsilon }{r}} \right) + 2\nu _0 \frac{{\Phi (t)}}{{r^2 }} =
0,\,\,\,\,r = R_{1,2} (t)
\]
\[
\frac{{d\,\Phi }}{{dt}}\ln \left( {\frac{{R_1 }}{{R_2 }}} \right)
= \int\limits_{R_2 (t)}^{R_1 (t)} {\left[ {\frac{{\upsilon ^2
}}{r} + 2\nu _0 \frac{\partial }{{\partial r}}\left(
{\frac{\upsilon }{r}} \right)} \right]} \,dr - \left( {2\nu \Phi -
\frac{{\Phi ^2 }}{2}} \right)\left[ {\frac{1}{{R_2^2 }} -
\frac{1}{{R_1^2 }}} \right];
\]
\[
\Phi (0) = \Phi _0;
\]
\[
\upsilon (r,0) = \upsilon _0 (r);
\]
The equality
\[ R_1^2 (t) - R_2^2 (t) = R_{10}^2  - R_{20}^2;
\]
is the conservation law of square.
The equality
\[ \int\limits_{R_2 (t)}^{R_1 (t)} {r^2 \upsilon
(r,t)\,dr = \int\limits_{R_{20} }^{R_{10} } {r^2 \upsilon
(r,0)\,dr} } \equiv \int\limits_{R_{20} }^{R_{10} } {r^2 \upsilon
_0 (r)\,dr};
\]
is the conservation law of impulse moment.

Let us write down the equation for temperature $T$:
$$
\rho \,C_p \left( {\frac{{\partial T}}{{\partial t}} +
u\frac{{\partial T}}{{\partial r}} + \frac{\upsilon
}{r}\frac{{\partial T}}{{\partial \theta }}} \right) = k\left[
{\frac{1}{r}\frac{\partial }{{\partial r}}\left( {r\frac{{\partial
T}}{{\partial r}}} \right) + \frac{1}{{r^2 }}\frac{{\partial ^2
T}}{{\partial \theta ^2 }}} \right] +{}$$
$${}+ \mu \left\{
{2\left( {\frac{{\partial u}}{{\partial r}}} \right)^2  + 2\left(
{\frac{u}{r}} \right)^2  + \left[ {r\frac{\partial }{{\partial
r}}\left( {\frac{\upsilon }{r}} \right)} \right]^2 } \right\};
$$
\[
\left. T \right|_{t = 0}  = T_0 (r,\theta ).
\]
Boundary conditions for $T$ will be specified additionally.

Let
\begin{eqnarray*}
 \tau  = \frac{\nu }{{R_{20} ^2 }}t;\qquad \eta  = \frac{{r^2  - R_2 ^2
(t)}}{{R_{20} ^2 }};\qquad \xi  = \frac{{R_2 ^2 (t)}}{{R_{20} ^2
}};\qquad
\\
u = \frac{{\nu \,\Psi }}{r};\qquad  \upsilon = \frac{\nu
}{{R_{20}^2 }}r\omega;\qquad \Psi  = \frac{\Phi }{\nu };\qquad a =
\frac{{R_{10}^2 }}{{R_{20}^2 }} - 1;
\\
\Theta = \frac{T}{{T_0 }};\qquad \Theta _0  =
T_0\qquad\qquad\qquad\qquad
\end{eqnarray*}
be a new coordinates then we have a problem:
\begin{equation}\label{Belmetsev11}
    \frac{{\partial \omega }}{{\partial \tau }} +%
 \frac{{2\Psi }}{{\xi  + \eta }}\omega  =%
  \\4(\xi  + \eta )\frac{{\partial ^2 \omega }}{{\partial \eta ^2 }} +%
   8\frac{{\partial \omega }}{{\partial \eta
}};
\end{equation}
\begin{equation}\label{Belmetsev12}
   \left. \omega  \right|_{\tau  = 0}  = \omega _0 (\eta ),\qquad 0 \le \eta  \le
   a;
\end{equation}
\begin{equation}\label{Belmetsev13}
   \frac{{\partial \omega }}{{\partial \eta }} + \varepsilon
\frac{\Psi }{{(\xi  + \eta )^2 }} = 0,\,\,\,\, \textrm{if $\eta =
0,\,a$;}\qquad \tau \ge 0;
\end{equation}
\begin{equation}\label{Belmetsev14}
   \frac{{d\omega }}{{d\tau }} = \frac{{a\Psi (\Psi  - 4)}}{{\xi
(\xi + a)\ln (1 + \frac{a}{\xi })}} + \frac{1}{{\ln (1 +
\frac{a}{\xi })}}\int\limits_0^a {\left( {\omega ^2  +
4\varepsilon \frac{{\partial \omega }}{{\partial \eta }}}
\right)\,d\eta };
\end{equation}
\begin{equation}\label{Belmetsev15}
   \frac{{d\xi }}{{d\tau }} = 2\Psi;\qquad \Psi (0) = \Psi _0;\qquad \xi (0) =
1;\qquad \varepsilon  = \frac{{\nu _0 }}{\nu };
\end{equation}
$$
 \frac{{\rho \,C_p \,R_{20}^2 \,\Theta _0 }}{{4\mu ^2 }}%
 \left( {\frac{{\partial \Theta }}{{\partial \tau }} +%
  \omega \frac{{\partial \Theta }}{{\partial \theta }}} \right)
  = {}
$$
$$
 {} = \frac{{k\,R_{20}^2 \,\Theta _0 }}{{4\mu ^3 }}%
 \left( {\frac{{\partial \Theta }}{{\partial \eta }} +%
 (\xi  + \eta )\frac{{\partial ^2 \Theta }}{{\partial \eta ^2 }} +%
 \frac{1}{{4\,(\xi  + \eta )}}\frac{{\partial ^2 \Theta }}{{\partial \theta ^2 }}} \right) + {}
$$
\begin{equation}
 \label{Belmetsev16}
 {} + \frac{{\Psi ^2 }}{{(\xi  + \eta )^2 }} + (\xi  + \eta )^2 (\frac{{\partial \omega }}{{\partial \eta }})^2
\end{equation}

It is easy to see that
\begin{equation} \label{Belmetsev17}
 \omega  = \frac{{4\varepsilon }}{{\xi  + \eta }};\quad \omega _0
 (\eta ) = \frac{{4\varepsilon }}{{1 + \eta }};\quad \Psi  = \Psi _0 =
 4;\quad \xi (\tau ) = 8\tau  + 1
\end{equation}
is an exact solution of the problem
\eqref{Belmetsev11}--\,\eqref{Belmetsev15}. Using
\eqref{Belmetsev16},\eqref{Belmetsev17}, and $\Theta = \Theta
(\tau (t),\eta (r,t))$, we get:
\begin{equation}\label{Belmetsev18}
 A\frac{{\partial \Theta }}{{\partial \tau }} = B\frac{\partial
}{{\partial \eta }}\left( {(8\tau  + \eta  + 1)\frac{{\partial
\Theta }}{{\partial \eta }}} \right) + \frac{{16\,(1 + \varepsilon
^2 )}}{{(8\tau  + \eta  + 1)^2 }}
\end{equation}
Here
\begin{equation}\label{Belmetsev19}
 A = \frac{{\rho \,C_p \,R_{20}^2 \,\Theta _0 }}{{4\mu ^2 }},\qquad B =
 \frac{{k\,R_{20}^2 \,\Theta _0 }}{{4\mu ^3 }},\qquad \varepsilon  = \frac{{\nu _0 }}{\nu }
\end{equation}
are dimensionless parameters.

Using methods of Group Analysis (see \cite{OVSYANNIKOV2B,
IBRAGIMOV1B, IBRAGIMOV2B, MELESHKO1B}), we obtain infinitesimal
operator:
\begin{equation} \label{Belmetsev20}
  \mathop X = \xi ^1 \frac{\partial
  }{{\partial \tau }} + \xi ^2 \frac{\partial }{{\partial \eta }} +
  \eta ^1 \frac{\partial }{{\partial \Theta }}
\end{equation}
with coefficients:
\begin{equation}
\label{Belmetsev21}
 \left\{
 \begin{array}{l}
  {\xi ^1 = {C_1 }{\tau ^2}/{2}  + C_2 \,\tau  + C_3;}
\\
  {\xi ^2 = C_1 \,\tau \,(4\tau  + \eta  + 1) + C_2 \,(\eta  + 1) -
8\,C_3;}
\\
  {\eta ^1 = \left( {C_4  - {C_1 }\left( {\tau + ({A}/{B})\eta} \right)/{2}} \right)\Theta +
 b_2(\tau,\eta);}
 \end{array}
 \right.
\end{equation}
Here $C_i$ are arbitrary real constants $(i = 1,2,3,4)$. A
function $b_2(\tau,\eta)$ should be a solution of following
equation:
$$
  A\frac{{\partial b_2 }}{{\partial \tau }} =
  B\,\left( {\frac{{\partial b_2 }}{{\partial \eta }} + (8\,\tau  +
  \eta  + 1)\frac{{\partial ^2 b_2 }}{{\partial \eta ^2 }}} \right)
  + {}
$$
\begin{equation}
\label{Belmetsev22}
  {} + \frac{{16\,(1 + \varepsilon ^2 )}}{{(8\,\tau  + \eta  + 1)^2
  }}\left( { - (C_2  + C_4 ) - \frac{{C_1 }}{2}\left( {\tau  -
  \frac{A}{B}\eta } \right)} \right)
\end{equation}


Supposing that $C_1 =0$, $C_2 =0$, $C_4 =0$, and
$b_2(\tau,\eta)=0$, we get:
$$
\mathop X = C_3 \frac{\partial }{{\partial \tau }} - 8\,C_3
\frac{\partial }{{\partial \eta }}
$$
$$
\left\{ {\begin{array}{*{20}l}
  {I_1  = 8\,\tau  + \eta  + 1}\\
  {I_2  = \Theta }\\
\end{array}} \right.
$$
Using the theorem for representation of invariant solutions
through invariants from \cite{OVSYANNIKOV2B}, we receive for
unknown function $\Theta$:
$$
\Theta  = I_2  = \varphi \,(I_1 ) = \varphi \,(8\,\tau  + \eta  +
1)
$$
Combining this with \eqref{Belmetsev18} we get the ordinary
differential equation:
$$
B\,I_1 \frac{{d^2 \varphi \,(I_1 )}}{{d\,I_1}^2} + (B -
8A)\frac{{d\,\varphi \,(I_1 )}}{{d\,I_1 }} + \frac{{16\,(1 +
\varepsilon ^2 )}}{{(I_1 )^2 }} = 0
$$
Solving this equation, and checking solution by input equation we
obtain the invariant solution of equation \eqref{Belmetsev18}:
\begin{equation} \label{Belmetsev23}
\Theta (\tau ,\eta ) = C_5  - \frac{{16\,(1 + \varepsilon ^2
)}}{{(8\,\tau  + \eta  + 1)(B + 8A)}}
\end{equation}

Let $C_1  = 0$, $C_2  = 1$, $C_4  =  - 2$, and function
$b_2(\tau,\eta)$ is the partial solution \eqref{Belmetsev23}:
$$
b_2(\tau ,\eta) = C_5  - \frac{{16\,(1 + \varepsilon ^2
)}}{{(8\,\tau  + \eta  + 1)(B + 8A)}}
$$
then for infinitesimal operator \eqref{Belmetsev20} (and for
invariants), we obtain:
$$
\mathop X = \left(\tau + C_3\right) \frac{\partial }{{\partial
\tau }} + \left(1 + \eta - 8\,C_3\right) \frac{\partial
}{{\partial \eta }} + \left( b_2(\tau ,\eta) - 2\Theta
\right)\frac{\partial }{{\partial \Theta }}
$$
$$
\left\{ {\begin{array}{*{20}l}
  {J_1  = (1 + \eta  - 8C_3)/(\tau  + C_3)} \\
  {J_2  = (\tau  + C_3 )\left( {(\tau  + C_3 )\Theta - \tau C_5/2 } \right) - \tau C_3 C_5 /2 +{}} \\
  {{}+ 16 \,\tau (\tau  + C_3 ) (1 + \varepsilon ^2 )/((8\tau  + \eta  + 1)(8A +
  B))}
\end{array}} \right.
$$
Using $J_1$, $J_2$, we have (in system of computer calculus Maple)
the invariant solution for equation \eqref{Belmetsev18}:
$$
\Theta (\tau ,\eta ) = K \left( {\frac{{A (1 + \eta  - 8 C_3 ) - B
(\tau  + C_3 )}}{{(\tau  + C_3 )^3 }}} \right)\,\times
$$
$$
 {} \times \exp \left\{ { - \frac{{A (1 + \eta  - 8 C_3 )}}{{B (\tau  + C_3
 )}}} \right\} \left( {\frac{{8 \tau  + \eta  + 1}}{{\tau  + C_3
 }}} \right)^{\frac{{8A}}{B}} + {}
$$
\begin{equation} \label{Belmetsev24}
 {} + \frac{{C_5 }}{2} - \frac{{16 (1 + \varepsilon ^2 )}}{{(8 \tau  + \eta  + 1)(B + 8A)}}, \\
\end{equation}
$$K = \frac{{8A\,K_1  + B\,(K_1  - K_2 )}}{{B\,(8A + B)}} =
const.$$ Here $K_1$, $K_2$ are constants. It is easy to see that
the expressions \eqref{Belmetsev23},~\eqref{Belmetsev24} are
solutions for equation \eqref{Belmetsev18}, and:
\begin{equation*}
\mathop {\lim }\limits_{\tau  \to  + \infty } \,\left( {\Theta
(\tau ,\eta )} \right) = \frac{{C_5 }}{2}.
\end{equation*}

Let
\begin{equation} \label{Belmetsev25}
\left\{ {\begin{array}{*{20}c}
   {\Theta _1 (\tau ) = \Theta (\tau ,a)} \hfill  \\
   {\Theta _2 (\tau ) = \Theta (\tau ,0)} \hfill  \\
\end{array}} \right.
\end{equation}
be a boundary conditions.

Using \eqref{Belmetsev24} and \eqref{Belmetsev25} we receive:

\begin{equation} \label{Belmetsev26}
\left\{ {\begin{array}{*{20}c}
   \begin{array}{l}
 \Theta _1 (\tau ) = K\,\left( {\frac{{A\,(1 + a - 8\,C_3 ) - B\,(\tau  + C_3 )}}{{(\tau  + C_3 )^3 }}} \right)\,\exp \,\left\{ { - \frac{{A\,(1 + a - 8\,C_3 )}}{{B\,(\tau  + C_3 )}}} \right\}\,\left( {\frac{{8\,\tau  + a + 1}}{{\tau  + C_3 }}} \right)^{\frac{{8A}}{B}}  +  \\
  + \frac{{C_5 }}{2} - \frac{{16\,(1 + \varepsilon ^2 )}}{{(8\,\tau  + a + 1)(B + 8A)}} \\
 \end{array} \hfill  \\
   \begin{array}{l}
 \Theta _2 (\tau ) = K\,\left( {\frac{{A\,(1 - 8\,C_3 ) - B\,(\tau  + C_3 )}}{{(\tau  + C_3 )^3 }}} \right)\,\exp \left\{ { - \frac{{A\,(1 - 8\,C_3 )}}{{B\,(\tau  + C_3 )}}} \right\}\,\left( {\frac{{8\,\tau  + 1}}{{\tau  + C_3 }}} \right)^{\frac{{8A}}{B}}  +  \\
  + \frac{{C_5 }}{2} - \frac{{16\,(1 + \varepsilon ^2 )}}{{(8\,\tau  + 1)(B + 8A)}} \\
 \end{array} \hfill  \\
\end{array}} \right.
\end{equation}

The difference of temperatures on boundaries at $\tau  = 0$ is
defined by the constant:

$$
\begin{array}{l}
 C = \Theta _1 (0) - \Theta _2 (0) = \frac{{16\,a\,(1 + \varepsilon ^2 )}}{{(1 + a)(8A + B)}} + \frac{K}{{C_3 ^3 }}\,\exp \left\{ { - \frac{{A\,(1 - 8\,C_3 )}}{{B\,C_3 }}} \right\}\, \times  \\
  \times \,\left( {\left( {A(1 + a - 8\,C_3 ) - B\,C_3 } \right)\,\exp \left\{ { - \frac{{A\,a}}{{B\,C_3 }}} \right\}\,\left( {\frac{{1 + a}}{{C_3 }}} \right)^{\frac{{8A}}{B}}  - \left( {A(1 - 8\,C_3 ) - B\,C_3 } \right)\,\left( {\frac{1}{{C_3 }}} \right)^{\frac{{8A}}{B}} } \right) \\
 \end{array}
$$

We have distribution of temperature $\Theta (\tau ,\eta )$ for
$\tau  = 0$:

\begin{equation} \label{Belmetsev27}
\begin{array}{l}
 \Theta (0,\eta ) = K\,\left( {\frac{{A\,(1 + \eta  - 8\,C_3 ) - B\,C_3 }}{{C_3 ^3 }}} \right)\,\exp \,\left\{ { - \frac{{A\,(1 + \eta  - 8\,C_3 )}}{{B\,C_3 }}} \right\}\,\left( {\frac{{\eta  + 1}}{{C_3 }}} \right)^{\frac{{8A}}{B}}  +  \\
  + \frac{{C_5 }}{2} - \frac{{16\,(1 + \varepsilon ^2 )}}{{(\eta  + 1)(B + 8A)}}, \\
 \end{array}
\end{equation}

Let us choose constants as follows:

$$
 C = 0,\qquad A = \frac{3}{4},\qquad B = 6,\qquad \varepsilon
= \frac{1}{2},\qquad a = 1,\qquad C_3  = \frac{1}{8},
$$
\begin{equation} \label{Belmetsev27}
K = \frac{{\frac{{10}}{9}C_3 ^4 }}{{4\,(8\,C_3  - 1)\,\exp \left\{
{1 - \frac{1}{{4\,C_3 }}} \right\} - (16\,C_3  - 1)\exp \left\{ {1
- \frac{1}{{8\,C_3 }}} \right\}}} =  - \frac{5}{{6^2 \,8^3 }}
\approx  - 0.00027127
\end{equation}
Values of temperature on boundaries at $\tau = 0$ with such choice
of constants are identical.

Considering values of constants \eqref{Belmetsev27} and
expressions for temperature on boundaries \eqref {Belmetsev26}, we
have the following problem:

\begin{equation} \label{Belmetsev28}
\frac{{\partial \Theta }}{{\partial \tau }} - \frac{\partial
}{{\partial \eta }}\left( {8\,(8\tau  + \eta  + 1)\frac{{\partial
\Theta }}{{\partial \eta }}} \right) - \frac{{80}}{{3\,(8\tau  +
\eta  + 1)^2 }} = 0
\end{equation}

\begin{equation} \label{Belmetsev29}
\left\{ {\begin{array}{*{20}c}
   {\left. {\frac{{\partial \Theta }}{{\partial \eta }}} \right|_{\,\eta  = 0}  = \frac{{5\,(1 + 16\,\tau )}}{{6\,(8\,\tau  + 1)^3 }}} \hfill  \\
   {\left. {\frac{{\partial \Theta }}{{\partial \eta }}} \right|_{\,\eta  = a}  = \left( {\frac{5}{{3\,(8\,\tau  + 2)^2 }}} \right) + \left( {\frac{5}{{6\,(8\,\tau  + 2)^3 }}} \right)\,\left( {\frac{{1 - (8\,\tau  + 1)(8\,\tau  + 3)}}{{(8\,\tau  + 1)^2 }}} \right)} \hfill  \\
\end{array}} \right.
\end{equation}

\begin{equation} \label{Belmetsev30}
\Theta (0,\eta ) = \frac{{C_5 }}{2} - \left( {\frac{5}{6}}
\right)\,\left( {(\eta ^2  - 1)\,\exp ( - \eta ) + \frac{2}{{(\eta
+ 1)}}} \right)
\end{equation}

Substituting values of constants \eqref {Belmetsev27} in
expression \eqref {Belmetsev24}, we receive the solution of the
problem \eqref {Belmetsev28}-\eqref {Belmetsev30}:

\begin{equation} \label{Belmetsev31}
\Theta (\tau ,\eta ) = \frac{{C_5 }}{2} - \left( {\frac{5}{6}}
\right)\,\left( {\frac{2}{{(8\,\tau  + \eta  + 1)}} + \left(
{\frac{{\eta ^2  - (8\,\tau  + 1)^2 }}{{(8\,\tau  + 1)^4 }}}
\right)\,\exp \left\{ { - \frac{\eta }{{8\,\tau  + 1}}} \right\}}
\right)
\end{equation}

Supposing temperature nonnegative, we get restriction on value of
a free constant: $C_5 \ge {5} / {3} $.






Considering \eqref{Belmetsev17}, we get the solution
\eqref{Belmetsev24} in $(t, r)$ co-ordinates:
$$
T(t,r) = T_0 K \left( {\frac{{A R_{20}^4 r^2 - (8A + B) (\nu
R_{20}^4 t + C_3 R_{20}^6)}}{{(\nu t + C_3 R_{20}^2)^3 }}}
\right)\,\times
$$
$$
 {} \times \exp \left\{ \frac{8A}{B} - {\frac{{A r^2}}{{B(\nu t + C_3 R_{20}^2)}}} \right\} \left( {\frac{r^2}{(\nu t + C_3 R_{20}^2)}} \right)^{\frac{{8A}}{B}} + {}
$$
$$
 {} + \frac{{C_5 T_0}}{2} - \frac{{16 T_0 R_{20}^2 (1 + \varepsilon ^2)}}{{r^2(B + 8A)}}. \\
$$

Also, we have:
$$
R_{20}^2(t) = 8 \nu t + R_{20}^2, \qquad R_{10}^2(t) = 8 \nu t +
R_{10}^2,
$$
$$
u = \frac{4\nu}{r}, \qquad \upsilon = \frac{4\nu_0}{r}.
$$

As a result, exact solutions are received, the general view of Lie
group for the equation (19) is  received, the asymptote is found,
and profiles for one of the solutions are constructed. These
results allows to build another solutions for a problem about
temperature distribution in the liquid ring.


\end{document}